\documentclass[twocolumn,superscriptaddress,pra]{revtex4}
\usepackage[dvips]{graphicx,color}
\usepackage{delarray}
\usepackage{amsmath}
\usepackage{bm}
\usepackage{ogonek}

\newcommand{\braket}[2]{\langle {#1} | {#2} \rangle}
\newcommand{\ket}[1]{| { #1} \rangle}
\newcommand{\bra}[1]{ \langle {#1}  |}

\begin{document}
\title{Optimal supplier of single-error-type entanglement via coherent-state transmission}

\author{Koji Azuma}
\email{koji.azuma.ez@hco.ntt.co.jp}
\affiliation{NTT Basic Research Laboratories, NTT Corporation, 3-1 Morinosato Wakamiya, Atsugi, Kanagawa 243-0198, Japan}
\affiliation{NTT Research Center for Theoretical Quantum Physics, NTT Corporation, 3-1 Morinosato-Wakamiya, Atsugi, Kanagawa 243-0198, Japan}

\author{Nobuyuki Imoto}
%\email{}
\affiliation{Institute for Photon Science and Technology, School of Science, University of Tokyo, Tokyo 113-0033, Japan}

\author{Masato Koashi}
%\email{}
\affiliation{Photon Science Center, University of Tokyo, Hongo, Bunkyo-ku, Tokyo 113-8656, Japan}

\
\date{\today}

%%%%%%%%%%%%%%%%%%%%%%%%%%%% abstract %%%%%%%%%%%%%%%%%%%%%%%%%%%%%
\begin{abstract}
Compared with entanglement with multiple types of noise, entanglement including only one type of error is a favorable fundamental resource not only for quantum communication but also for distributed quantum computation. 
We consider protocol that presents single-error-type entanglement for distant qubits via coherent-state transmission over a lossy channel. This protocol is regarded as a subroutine to serve entanglement for larger protocol to yield a final output, such as ebits or pbits. In this paper, we provide a subroutine protocol which achieves the {\it global} optimal for typical jointly convex yield functions monotonically non-decreasing with respect to the singlet fraction, such as an arbitrary convex function of a singlet fraction and two-way distillable entanglement/key. Entanglement generation based on remote non-destructive parity measurement protocol [K. Azuma, H. Takeda, M. Koashi, and N. Imoto, Phys. Rev. A {\bf 85}, 062309 (2012)]  is identified as such an optimal subroutine.
\end{abstract}
\maketitle

\section{Introduction}

Generating entanglement between distant qubits is a fundamental building block not only for quantum communication but also for distributed quantum computation. In quantum key distribution (QKD), private bits (pbits) are distilled from the (virtually) generated entangled states through error correction and privacy amplification \cite{M01,LC99,SP00,K09,R05}, while Bell pairs (ebits) are through entanglement distillation \cite{B96,B96a,D96} for more general scenarios such as quantum teleportation \cite{B93} and controlled-not operations \cite{G99,CLP01,EJPP00} for spatially distant qubits/chips in distributed fault-tolerant quantum computation \cite{FSG09,FYKI12,LB12,NLS13,M14,N14}. 
If the entanglement generation protocol is run in parallel between nearest-neighboring nodes in a quantum network, ebits/pbits are served for arbitrary clients efficiently \cite{B98,D99,AK17}, through entanglement distillation and entanglement swapping \cite{Z93}. Therefore, entanglement generation protocol is regarded in general as a subroutine to serve entanglement for larger protocol to yield a final output, such as ebits or pbits.

Entanglement generation protocol is normally based on transmission of flying bosonic systems, such as photons, over a communication channel, such as an optical fiber, a mode in free space, or a microwave transmission line. The dominant noise in the channel is the photon loss.  Recently, upper bounds on the two-way quantum/private capacity of a point-to-point pure-loss bosonic channel are derived \cite{TGW14,PLOB17}. These bounds show that the performance of existing point-to-point entanglement generation protocol \cite{L08,M08,A09,AK12,M10,M12} or QKD protocol \cite{BB84,H03,LMC05,W05,B92,GG02,K04,SYK14,T14} based on the transmission of polarized (or time-bin) single-photon states, Fock states, or coherent states (or cat states) over a pure-loss channel has no scaling gap with the upper bounds. 
Entanglement generation protocol \cite{D01,C06,ATKI10} or QKD protocol \cite{AKB14,PRML14,ATM15,LYDS18,C19,CAL19,MZZ18,LL18,LLKL16,Ro18,Ro19,MSK19,LNAKCR21,WYH18} working with an intermediate node between communicators also has no scaling gap with upper bounds \cite{AK17,AML16,P19,R18} on the quantum/private capacity of the corresponding quantum network (see a review article \cite{A21}). 
Besides, if such entanglement generation protocol with no scaling gap with the bounds is combined with the optimal entanglement distillation and entanglement swapping to present ebits/pbits to clients over a pure-loss bosonic channel network, its performance has no scaling gap even with the quantum/private capacity of the network, irrespective of its topology \cite{AK17}. These facts suggest that there is not much room to improve further existing protocols in terms of scaling. In other words, it is rather important in practice to design a protocol by considering a balance between easiness of the implementation and its specific performance.

Especially, protocol based on coherent-state encoding is an example of protocol with such a good balance. In B92 QKD protocol \cite{B92}, a bit is encoded into phases of a coherent state, and it is sent from a sender, Alice, to a receiver, Bob, directly through an optical channel. In the twin-field QKD protocol \cite{LYDS18,C19,CAL19,LL18,WYH18,MSK19,LNAKCR21,MZZ18}, Alice and Bob send coherent states with information of bits to an intermediate node, called Claire, which is supposed to perform a Bell measurement based on single-photon interference.
Entanglement generation protocol \cite{L06,L08,M08,A09,AK12,M10,C06,ATKI10} is also based on coherent-state encoding to generate entanglement between Alice's qubit and Bob's qubit, as such an encoding can be established through a dispersive Jaynes-Cummings Hamiltonian between a coherent state and a matter qubit, such as a superconducting qubit, a quantum dot, a single ion, a nitrogen-vacancy center in a diamond, or a single atom. Therefore, protocols based on coherent-state encoding constitute an important category as practical entanglement generation and QKD.

In this paper, we consider protocol that presents entanglement with only one type of error (such as a phase error) for distant qubits via coherent-state transmission over a lossy channel, as well as local operations and classical communication (LOCC). On regarding this as a subroutine to serve single-error-type entanglement for larger protocol to yield a final output, we identify a protocol which achieves the {\it global} optimal for typical jointly convex yield functions monotonically non-decreasing with respect to the singlet fraction \cite{H}, such as an arbitrary convex function of a singlet fraction and two-way distillable entanglement/key. 
In particular, entanglement generation protocol based on remote non-destructive parity measurement (RNPM) protocol \cite{A09,ATKI10} is identified as such an optimal subroutine.

This paper is organized as follows.
In Sec.~\ref{se:yield}, we define the yield function for single-error-type entanglement and show its several properties. 
We consider point-to-point protocol in Sec.~\ref{se:2} and three-party protocol working with the help of an intermediate node in Sec.~\ref{se:3}. 
Section~\ref{se:4} concludes this paper.

\section{Single-error-type entanglement and the yield based on it}\label{se:yield}

In this paper, we consider entanglement generation protocols which provide single-error-type entanglement $\hat{\tau}^{AB}$ for distant qubits $A$ and $B$, i.e., a state in the Hilbert subspace spanned by two orthogonal Bell states, with the free use of LOCC. 
Since two orthogonal Bell states can be transformed into $\ket{\Phi^+}_{AB}:=(\ket{00}_{AB}+\ket{11}_{AB})/\sqrt{2}$ and $\ket{\Phi^-}_{AB}:=(\ket{00}_{AB}-\ket{11}_{AB})/\sqrt{2}$ via a local unitary operation \cite{B96}, we can assume that the state is described as
\begin{multline}
\hat{\tau}^{AB}(\zeta,\chi,\upsilon)=  \frac{1+\zeta}{2}  \ket{ \Phi^{+}}\bra{ \Phi^{+}}_{AB} + \frac{1-\zeta}{2}  \ket{ \Phi^{-}}\bra{ \Phi^{-}}_{AB}  \\
 + \frac{\chi-i\upsilon}{2} \ket{ \Phi^{+}}\bra{ \Phi^{-}}_{AB} + \frac{\chi+i\upsilon}{2} \ket{ \Phi^{-}}\bra{ \Phi^{+}}_{AB}   \label{eq:sing}
\end{multline}
with three real parameters $\zeta$, $\chi$, and $\upsilon$ satisfying $\zeta^2+\chi^2+\upsilon^2 \le 1$. By noting that $\hat{X}^A\otimes \hat{X}^B \ket{\Phi^\pm}_{AB} =\pm \ket{\Phi^\pm}_{AB} $ and $\hat{Z}^A\ket{\Phi^+}_{AB} =\ket{\Phi^-}_{AB}$ and that the application of a unitary operation $e^{-i \theta \hat{Z}^A/2}$ is closed in the Hilbert subspace spanned by $\{\ket{\Phi^\pm}_{AB}\}$, where $\hat{X}^A:=\ket{0}\bra{1}_A+\ket{1}\bra{0}_A$ and $\hat{Z}^A:=\ket{0}\bra{0}_A-\ket{1}\bra{1}_A$, the single-error-type entanglement $\hat{\tau}^{AB}$ can always be transformed into a standard form 
\begin{align}
\hat{\gamma}^{AB}(z,x):=&  \frac{1+z}{2}  \ket{ \Phi^{+}}\bra{ \Phi^{+}}_{AB} + \frac{1-z}{2}  \ket{ \Phi^{-}}\bra{ \Phi^{-}}_{AB}  \nonumber \\
& + \frac{x}{2}( \ket{ \Phi^{+}}\bra{ \Phi^{-}}_{AB} + \ket{ \Phi^{-}}\bra{ \Phi^{+}}_{AB})   \nonumber \\
=&  \frac{1+x}{2}  \ket{00}\bra{00}_{AB} + \frac{1-x}{2}  \ket{11}\bra{11}_{AB}  \nonumber \\
& + \frac{z}{2}( \ket{00}\bra{ 11}_{AB} + \ket{11}\bra{00}_{AB}), \label{eq:standard}
\end{align}
via a local unitary operation,
where 
\begin{equation}
\begin{split}
x=&|\chi|,\\
z=&\sqrt{\zeta^2+\upsilon^2} \label{eq:zx}
\end{split}
\end{equation}
are nonnegative parameters satisfying $x^2+z^2\le 1$.
Note that $z$ is related with the {\it singlet fraction} $F$ of $\hat{\tau}^{AB}$ \cite{BHHH00}, defined by $F:=\max_{\hat{U}^A\otimes \hat{V}^B} {}_{AB}\bra{\Phi^+} (\hat{U}^A\otimes \hat{V}^B) \hat{\tau}^{AB} (\hat{U}^{A\dag} \otimes \hat{V}^{B\dag}) \ket{\Phi^+}_{AB}$ with unitary operators $\hat{U}^A$ and $\hat{V}^B$, as
\begin{equation}
z=2F-1.
\end{equation}

We consider a scenario where a single-error-type entangled state $\hat{\tau}^{AB}$ generated through an entanglement generation protocol can be used as an input for a subsequent protocol such as entanglement distillation, secret-key distillation, entanglement swapping \cite{Z93}, or their combination. In particular, we assume that the subsequent protocol accepts only the standard form $\hat{\gamma}^{AB}(z,x)$ and its yield $Y$ is a function of $z$ and $x$, i.e., $Y=Y\left( \hat{\gamma}^{AB}(z,x)\right)=Y(z,x)$. 
Using the yield function $Y(z,x)$ of a subsequent protocol as a reference, we may define a measure of entanglement in  general single-error-type state $ \hat{\tau}^{AB}(\zeta,\chi,\upsilon) $, which we also denote by $Y$ as $Y=Y\left( \hat{\tau}^{AB}(\zeta,\chi,\upsilon)\right)=Y(\sqrt{\zeta^2+\upsilon^2},|\chi|)$.

For $Y(\hat{\tau}^{AB})$ to be a proper measure, the yield function $Y(z,x)$ must satisfy several properties as follows. 
Since $Y(\hat{\tau}^{AB})$ should be zero for any separable state $\hat{\tau}^{AB}$,
the yield $Y$ is zero for separable states $\hat{\gamma}^{AB}(0,x)$, i.e.,
\begin{equation}
Y(0,x) =0. \label{eq:sep}
\end{equation}
From the monotonicity of $Y(\hat{\tau}^{AB})$ under LOCC as an entanglement measure, 
if Alice and Bob can deterministically convert a state $\hat{\gamma}^{AB}(z,x)$ to another state $\hat{\gamma}^{AB}(z',x')$ by LOCC, 
entanglement in $\hat{\gamma}^{AB}(z,x)$ is no smaller than that in $\hat{\gamma}^{AB}(z',x')$, namely,
\begin{equation}
Y(z,x) \ge Y(z',x').
\end{equation}
For example, if Alice inputs a qubit pair $AB$ in a state $\hat{\gamma}^{AB}(z,x)$ into a phase-flip channel 
\begin{equation}
\Lambda^A_v(\hat{\rho}) := \frac{1+v}{2} \hat{\rho} + \frac{1-v}{2} \hat{Z}^A \hat{\rho} \hat{Z}^A \label{eq:phase-flip-ch}
\end{equation}
with $0\le v \le 1$, the state of the qubit pair becomes $\hat{\gamma}^{AB}(vz,x)$, and thus, 
\begin{equation}
Y(z,x) \ge Y(vz,x), \label{eq:phase-flip-for-sing}
\end{equation}
implying monotonically non-decreasing of $Y(z,x)$ with respect to $z$.
Similarly, since Alice and Bob can convert state $\hat{\gamma}^{AB}(z,x)$ into state $\hat{\gamma}^{AB}(z,vx)$ by inputting the qubit pair into a channel 
\begin{equation}
{\cal E}^{AB}_v(\hat{\rho}) := \frac{1+v}{2} \hat{\rho} + \frac{1-v}{2} (\hat{X}^A \otimes \hat{X}^B ) \hat{\rho}(\hat{X}^A \otimes \hat{X}^B )
\end{equation}
with $0\le v \le 1$, we have
\begin{equation}
Y(z,x) \ge Y(z,vx), \label{eq:mono-x}
\end{equation}
implying monotonically non-decreasing of $Y(z,x)$ over $x$. 

In this paper, we impose two assumptions on the function $Y(z,x)$. 
That is to say, the results derived in the subsequent sections are true for any yield function $Y(z,x)$ as long as it satisfies those assumptions. The first assumption is that $Y(z,x)$ is a jointly convex function. This means that $Y$ is also a convex function over single-error-type states $\hat{\tau}^{AB}(\zeta,\chi,\upsilon)$, which can be seen as follows.
Consider a convex mixture of a single-error-type entangled state $\hat{\tau}^{AB}(\zeta',\chi',\upsilon')$ and a single-error-type entangled state $\hat{\tau}^{AB}(\zeta'',\chi'',\upsilon'')$ and denote it by a single-error-type state $\hat{\tau}^{AB}(\zeta,\chi,\upsilon)$. Then,
\begin{equation}
(\zeta,\chi,\upsilon)=p (\zeta',\chi',\upsilon') + (1-p) (\zeta'',\chi'',\upsilon'')
\end{equation}
holds for $0\le p\le 1$. The measure for the mixture $\hat{\tau}^{AB}(\zeta,\chi,\upsilon)$ then satisfies
\begin{multline}
Y(\hat{\tau}^{AB}(\zeta,\chi,\upsilon))=Y(\sqrt{\zeta^2+\upsilon^2},|\chi|)\\
\le Y(p\sqrt{\zeta'^2+\upsilon'^2}+(1-p)\sqrt{\zeta''^2+\upsilon''^2},p|\chi'|+(1-p)|\chi''|)\\
\le pY(\sqrt{\zeta'^2+\upsilon'^2},|\chi'|) +(1-p)Y(\sqrt{\zeta''^2+\upsilon''^2},|\chi''|)\\
=pY(\hat{\tau}^{AB}(\zeta',\chi',\upsilon')) +(1-p)Y(\hat{\tau}^{AB}(\zeta'',\chi'',\upsilon'')), \label{eq:conv}
\end{multline}
where we used the convexity of norms and the monotonicity of Eqs.~(\ref{eq:phase-flip-for-sing}) and (\ref{eq:mono-x}) to have the first inequality, and we used the joint convexity of the yield $Y(z,x)$ to have the second inequality.

The second assumption we make is an inequality 
\begin{equation}
Y(vz,\sqrt{1-z^2})\le z Y(v,0) \label{eq:mr}
\end{equation}
for any $0\le v\le 1$ and $0\le z\le 1$. The left-hand side of this inequality corresponds to the case where the output of the phase-flip channel  $\Lambda_v^A$ with the input of a pure state $\hat{\gamma}^{AB}(z,\sqrt{1-z^2})$ is sent to the subsequent protocol. On the other hand, the right-hand side corresponds to the case where a maximally entangled state $\hat{\gamma}^{AB}(1,0)$ given with probability $z$ is input to the phase-flip channel  $\Lambda_v^A$, followed by being sent to the subsequent protocol. The inequality (\ref{eq:mr}) requires the latter case to give an average overall yield to be no smaller than the former. Typical yield functions satisfy the inequality (\ref{eq:mr}), as inferred from the following examples.

As an example, let us consider a subsequent protocol whose yield function $Y$ depends only on the singlet fraction of $\hat{\gamma}^{AB}(z,x)$, i.e., $Y(z,x)=Y(z)$ for any $x$ and is convex over $z$. In this case, the convexity of $Y(z)$ and Eq.~(\ref{eq:sep}) imply
\begin{equation}
p Y(z) \ge Y(pz) \label{eq:mixsepg}
\end{equation}
for any $0\le p \le1$, leading to Eq.~(\ref{eq:mr}).
For instance, when we execute a quantum repeater protocol composed of single-error-type entanglement generation, the recurrence method \cite{B96,D96,B96a}, and the entanglement swapping,
it is conventional to start by converting initial entangled states $\hat{\gamma}^{AB}(z,x)$ between neighboring repeater stations into a Bell-diagonal one, 
$\hat{\gamma}^{AB}(z,0)$. Then, the yield function of the overall protocol would naturally depend only on the singlet fraction of the initial state, namely, $Y(z)$.

Another example of the yield function having convexity and satisfying Eq.~(\ref{eq:mr}) is the distillable entanglement $E_{\rm D}$.
The analytic formula of the distillable entanglement for general mixed states has not yet been found, 
but, it has been derived for maximally correlated states $\hat{\rho}^{AB}:=\sum_{i,j}a_{ij} \ket{ii}\bra{jj}_{AB}$ \cite{H,R99}.
The formula is described by
\begin{equation}
E_{\rm D} (\hat{\rho}^{AB}) = S(\hat{\rho}^{A}) -S(\hat{\rho}^{AB}),
\end{equation}
where $\hat{\rho}^{A}:={\rm Tr}_B [\hat{\rho}^{AB} ]$ and $S$ is the von Neumann entropy defined by $S(\hat{\rho}):=-{\rm Tr}[ \hat{\rho}  \log_2 \hat{\rho}]$. This quantity coincides with the two-way distillable key, in this case \cite{R99,HHHO05,HHHO09}.
Since the single-error-type entangled states $\hat{\gamma}^{AB} (z,x)$ are examples of the maximally correlated states, 
the distillable entanglement for $\hat{\gamma}^{AB} (z,x)$ is
\begin{equation}
E_{\rm D} (\hat{\gamma}^{AB} (z,x)) = h\left(\frac{1+x }{2} \right) - h\left( \frac{1+\sqrt{z^2 +x^2}}{2} \right), \label{eq:E_D}
\end{equation}
where $h$ is the binary entropy function $h(x):=-x \log_2 x -(1-x)\log_2(1-x)$.
A direct calculation shows that $E_{\rm D}$ is convex over $(z,x)$ and also satisfies Eq.~(\ref{eq:mr}).

\section{Tight bound on single-error-type entanglement generation}\label{se:2}

In this section, we derive a tight bound on entanglement generation protocols that are based on coherent-state transmission from a sender to a receiver, followed by arbitrary LOCC operations.
We start by defining the protocols and their yield as the measure of the performance (Sec.~\ref{se:2.1}).
In Sec.~\ref{se:2.2},
instead of considering an LOCC protocol that is generally complex, we consider separable operations and show the requisites for producing single-error-type entanglement.
In Sec.~\ref{se:2.3}, we derive an upper bound on the yield that could be given by the separable operations. Finally, in Sec.~\ref{se:2.4}, we show that a protocol of Ref.~\cite{A09} achieves the upper bound.

\subsection{Single-error-type entanglement generation and the measure of its performance}\label{se:2.1}

\begin{figure}[b]
    \includegraphics[keepaspectratio=true,height=75mm]{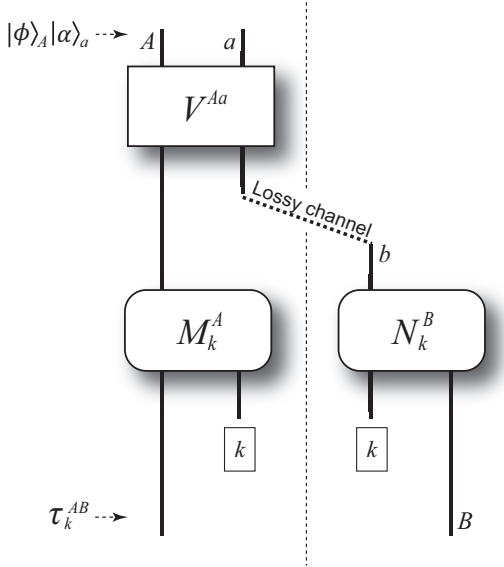}
  \caption{Scenario of single-error-type entanglement generation. $\ket{\phi}_A:=\sum_{j=0,1} \sqrt{q_j} e^{i \Theta_j} \ket{j}_A$. An outcome $k$ corresponds to the application of a separable operator $\hat{M}_k^A \otimes \hat{N}_k^B$. If the final entanglement $\hat{\tau}_k^{AB}$ includes only one type of error,
Alice and Bob may declare the success of the protocol.
   }
  \label{fig1a}
\end{figure}

Let us define the family of single-error-type entanglement generation protocols considered in this paper.
Suppose that separated parties called Alice and Bob have qubits $A$ and $B$, respectively, 
and their goal is to make the qubits $AB$ in a single-error-type entangled state.
In general, a protocol is described as follows (Fig.~\ref{fig1a}):
(i) Alice prepares qubit $A$ in her desired state $\ket{\phi}_A= \sum_{j=0,1} \sqrt{q_j} e^{i \Theta_j}  \ket{j}_A $ with real parameters $\Theta_j$, $q_j > 0$, and $\sum_j q_j=1$, and she makes it interact with a pulse in a coherent state $\ket{\alpha}_a=e^{-|\alpha|^2/2} e^{\alpha \hat{a}^\dag} \ket{0}_a$ via a unitary operation $\hat{V}^{Aa}$ defined by
\begin{equation}
\begin{split}
\hat{V}^{Aa} \ket{0}_A \ket{\alpha}_a &= \ket{0}_A \ket{\alpha_0}_a, \\
\hat{V}^{Aa} \ket{1}_A \ket{\alpha}_a &= \ket{1}_A \ket{\alpha_1}_a, \label{eq:V}
\end{split}
\end{equation}
where the possible output states $\{ \ket{\alpha_j}_a \}_{j=0,1}$ are also coherent states.
(ii) Alice sends the pulse $a$ to Bob, through a lossy channel described by an isometry
\begin{equation}
\hat{L}^{a\to bE} \ket{\alpha}_a = \ket{ \sqrt{T} \alpha}_b \ket{\sqrt{1-T} \alpha}_{E},
\end{equation}
where $0<T<1$ is the transmittance of the channel and system $E$ is the environment.
At this point, Alice and Bob share a quantum state described by
\begin{equation}
\ket{\psi}_{AbE} = \sum_{j=0,1} \sqrt{q_j} e^{i \Theta_j}  \ket{j}_A \ket{\sqrt{T}\alpha_j}_b \ket{\sqrt{1-T} \alpha_j}_E. \label{eq:initial}
\end{equation}
(iii)  
Alice and Bob manipulate system $Ab$ through LOCC, and may declare outcome $k$ with probability $p_k$ to herald the success of the generation of qubits $AB$ in a single-error-type entangled state $ \hat{\tau}_k^{AB}$.
The correction of the success events $k$ is denoted by ${\cal S}$.
We assume that for successful events with $k \in {\cal S}$, the state $\hat{\tau}_k^{AB}$ is given by the standard form in Eq. (\ref{eq:standard}), namely,
\begin{equation}
\hat{\tau}_k^{AB}=\hat{\gamma}^{AB} (z_k, x_k)  \label{eq:output-1}
\end{equation}
with $z_k>0$ and $x_k\ge 0$.

Let us define a method for evaluating single-error-type entanglement generation protocols. Typically,
following such an entanglement generation, a subsequent protocol that works with the obtained entanglement $\hat{\tau}_k^{AB}$, such as entanglement distillation, secret-key distillation, or entanglement swapping \cite{Z93}, is executed.
This implies that the value of the entanglement generation cannot be determined by itself, namely it depends on the protocol to be performed after the entanglement generation.
In this paper, using the yield function $Y$ defined in Sec.~\ref{se:yield}, 
the performance of the overall protocol is evaluated by the average overall yield $\bar{Y}$, which is defined by
\begin{align}
\bar{Y}&:=\sum_{k \in {\cal S} } p_k Y(\hat{\tau}_k^{AB})=\sum_{k \in {\cal S} } p_k Y\left(\hat{\gamma}^{AB}(z_k,x_k)\right) \nonumber\\
&=\sum_{k \in {\cal S} } p_k Y(z_k,x_k). \label{eq:performance-1}
\end{align}

\subsection{Requisites for separable operations}\label{se:2.2}

We start by considering the description of the LOCC in step (iii) of the protocol in Sec.~\ref{se:2.1}. 
In general, it is known that any LOCC operation can be described by a separable operation $\{\hat{M}_{\kappa}^{A} \otimes \hat{N}_{\kappa}^B \}$
(although the converse is not true, that is, there are separable operations \cite{B99,K07,CD09,KANI13} that are not implementable by LOCC).
Here $\kappa$ stands for the record of all the communication between Alice and Bob. The definition of the protocol in Sec.~\ref{se:2.1} allows the possibility of discarding part of the record, in which case 
the output state $\hat{\tau}_k^{AB}$ in step (iii) is a probabilistic mixture $\sum_\kappa q_{\kappa|k} \hat{\rho}^{AB}_{\kappa}$ over the output states $\{ \hat{\rho}^{AB}_{\kappa}\}_\kappa$ for various values of $\kappa$. Since $\hat{\tau}_k^{AB}$ is a single-error-type state in the form of Eq.~(\ref{eq:sing}), all the states $\{\hat{\rho}^{AB}_{\kappa}\}_\kappa$ should also be such single-error-type states. Then, due to the assumed convexity of the yield function $Y$ in Eq.~(\ref{eq:conv}), the optimum value of the average overall yield $\bar{Y}$ is always achieved by maintaining all the record. Hence we here assume that the state $\hat{\tau}_k^{AB}$ obtained in step (iii) is written by a single term as
\begin{equation}  
\hat{\tau}_k^{AB}= \frac{1}{p_k} (\hat{M}_k^A \otimes \hat{N}_k^B) {\rm Tr}_E(\ket{\psi} \bra{\psi}_{A bE})(\hat{M}_k^A \otimes \hat{N}_k^B)^\dag 
\label{eq:tau_k_in}
\end{equation}
with separable operators $\{\hat{M}_k^{A} \otimes \hat{N}_k^B \}$ satisfying
\begin{equation}
\sum_{k \in {\cal S}}\hat{M}_k^{A\dag} \hat{M}_k^A \otimes \hat{N}_k^{B\dag} \hat{N}_k^B \le \hat{1}^{AB}, \label{eq:completeness-rel}
\end{equation}
where $\hat{M}_k^A$ is an operator on the qubit $A$ while operator $\hat{N}_k^B$ maps state vectors for the system $b$ to those for the qubit $B$. Since $\hat{\tau}_k^{AB}$ with $k\in {\cal S}$ is entangled by definition, the ranks of operators $\hat{M}_k^A$ and $\hat{N}_k^B$ are 2 for any $k\in {\cal S}$.

We rewrite the state of Eq.~(\ref{eq:initial}) as
\begin{equation}
 \ket{\psi}_{A b E} = \sum_{j=0,1} \sqrt{q_j} e^{i \Theta_j}  \ket{j}_A \ket{u_j}_{b}
  \ket{v_j}_E 
\label{eq:abe}
\end{equation}
with $0 < q_0 < 1 $, $q_0+q_1=1$, and 
\begin{equation}
 1>|\braket{u_1}{u_0}|^{1-T} = |\braket{v_1}{v_0}|^T>0
 \label{tra}
\end{equation}
from a property $|\braket{\sqrt{T}\alpha_1}{\sqrt{T}\alpha_0}|=|\braket{\alpha_1}{\alpha_0}|^T$ of coherent states $\{\ket{\alpha_j}\}_{j=0,1}$.
From Eqs.~(\ref{eq:abe}) and (\ref{eq:phase-flip-ch}), we have a simplified representation \cite{A09},
\begin{equation}
{\rm Tr}_{E} [ \ket{\psi}\bra{\psi}_{A b E} ] = \Lambda^A_{|\braket{v_1}{v_0}|}(\ket{\psi'} \bra{\psi'}_{A b}), 
\end{equation}
where 
\begin{equation}
\ket{\psi'}_{A b}:= \sum_{j=0,1}  \sqrt{q_j} e^{i \Theta_j+ i (-1)^j  \varphi } \ket{j}_A \ket{u_j}_{b} \label{eq:psi'} 
\end{equation}
with $2 \varphi:=\arg [\braket{v_1}{v_0}]$.
Thus, Eq.~(\ref{eq:tau_k_in}) is rewritten as
\begin{equation}  
\hat{\tau}_k^{AB}= \frac{1}{p_k} (\hat{M}_k^A \otimes \hat{N}_k^B) \Lambda^A_{|\braket{u_1}{u_0}|}(\ket{\psi'} \bra{\psi'}_{A b})(\hat{M}_k^A \otimes \hat{N}_k^B)^\dag.
\label{eq:tau_k}
\end{equation}

\begin{figure}[b]
    \includegraphics[keepaspectratio=true,height=55mm]{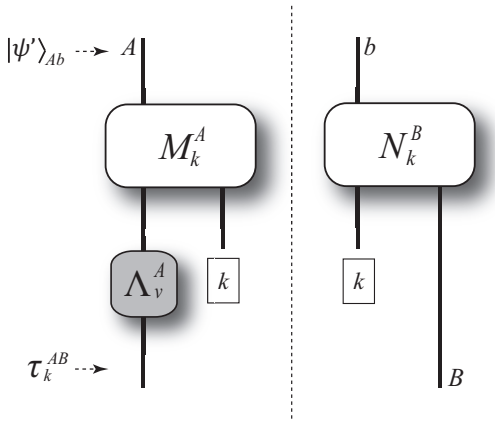}
  \caption{An imaginary protocol equivalent to the real protocol in Fig.~1. $\ket{\psi'}_{Ab}:=\sum_{j=0,1} \sqrt{q_j} e^{i \Theta_j+ i (-1)^j \varphi} \ket{j}_{A}\ket{u_j}_b$, where $\varphi:=\arg[\braket{v_1}{v_0}]/2$. Channel $a \to b$ becomes ideal at the expense of the application of a phase-flip channel $\Lambda^A_v$ with $v=|\braket{v_1}{v_0}|$. 
   }
  \label{fig2a}
\end{figure}

Let us consider requisites for $\{\hat{M}_k^A\otimes \hat{N}_k^B \}_{k\in {\cal S}}$, stemming from the assumption that $\hat{\tau}_k^{AB}$ is in the standard form of Eq.~(\ref{eq:standard}).
Since Eq.~(\ref{eq:standard}) implies ${}_{AB}\bra{01} \hat{\tau}_k^{AB} \ket{01}_{AB}=0$,
the separable operator $\hat{M}_k^A \otimes \hat{N}_k^B$ must satisfy
\begin{equation}
0={}_{AB}\bra{01} \hat{M}_k^A \Lambda^A_{|\braket{v_1}{v_0}|}( \hat{\sigma}_k^{AB} ) \hat{M}_k^{A \dag} \ket{01}_{AB}, \label{eq:single}
\end{equation}
where 
\begin{equation}
\hat{\sigma}_k^{AB}:=\hat{N}_k^B ( \ket{\psi'} \bra{\psi'}_{Ab}) \hat{N}_k^{B \dag}. \label{eq:defsigma}
\end{equation}
From $0<|\braket{v_1}{v_0}|<1$ and the positivity of $\hat{\sigma}_k^{AB}$, we have
\begin{multline}
\sqrt{\hat{\sigma}_k^{AB}} \ket{1}_B (\hat{M}_k^{A \dag} \ket{0}_A) =\sqrt{\hat{\sigma}_k^{AB}} \ket{1}_B ( \hat{Z}^A \hat{M}_k^{A\dag}  \ket{0}_A )=0. \label{eq:ld}
\end{multline}
If $\hat{M}_k^{A \dag} \ket{0}_A $ and $\hat{Z}^A \hat{M}_k^{A \dag} \ket{0}_A$ were linearly independent, $\sqrt{\hat{\sigma}_k^{AB}} \ket{1}_ B=0$, which would imply that $\hat{\sigma}_k^{AB}$ is a separable state. This would, in turn, mean the separability of $\hat{\tau}_k^{AB}$.
To avoid this contradiction, $\hat{M}_k^{A \dag} \ket{0}_A $ and $\hat{Z}^A \hat{M}_k^{A\dag} \ket{0}_A$ must be linearly dependent, which implies that the state $\hat{M}_k^{A \dag} \ket{0}_{A } $ is an eigenstate of $\hat{Z}^A$.
Similarly, 
from ${}_{AB}\bra{10} \hat{\tau}_k^{AB} \ket{10}_{AB}=0$, i.e.,
\begin{equation}
0={}_{AB}\bra{10} \hat{M}_k^A \Lambda^A_{|\braket{v_1}{v_0}|}( \hat{\sigma}_k^{AB} ) \hat{M}_k^{A \dag} \ket{10}_{AB}, \label{eq:single'}
\end{equation}
we have
\begin{equation}
\sqrt{\hat{\sigma}_k^{AB}} \ket{0}_B (\hat{M}_k^{A \dag} \ket{1}_A) =\sqrt{\hat{\sigma}_k^{AB}} \ket{0}_B ( \hat{Z}^A \hat{M}_k^{A\dag}  \ket{1}_A )=0, \label{eq:ld'}
\end{equation}
meaning that the state $\hat{M}_k^{A \dag} \ket{1}_A $ must also be an eigenstate of $\hat{Z}^A$.
Combined with ${\rm rank}(M_k^{A \dag})=2 $, these conclude that $\hat{M}_k^{A\dag} \ket{0}_A $ and $\hat{M}_k^{A\dag} \ket{1}_A $ are different eigenstates of $\hat{Z}^A$.
By letting $k\in {\cal S}_l$ for $l=0,1$ denote the subset of outcomes $k\in S$ such that $\hat{M}_k^{A\dag} \ket{0}_A \propto (\hat{X}^A)^l \ket{0}_A $ and $\hat{M}_k^{A\dag} \ket{1}_A \propto (\hat{X}^A)^l \ket{1}_A $, where ${\cal S}={\cal S}_0 \cup {\cal S}_1$ with ${\cal S}_0\cap {\cal S}_1 = \emptyset$, this implies
\begin{equation}
\hat{M}_k^A =  
\begin{cases}
m_k^{0} \ket{0}\bra{0}_A +m_k^{1} \ket{1}\bra{1}_A  & (k\in {\cal S}_0) ,\\
m_k^{0} \ket{1}\bra{0}_A +m_k^{1} \ket{0}\bra{1}_A  & (k\in {\cal S}_1),
\end{cases}
\label{eq:m-k}
\end{equation}
with nonzero $m_k^{j}$.
This equation shows that $\hat{M}_k^A$ commutes with the phase-flip channel $\Lambda^A_{|\braket{v_1}{v_0}|}$.
Hence, the considered protocol is simulatable by a protocol of Fig.~\ref{fig2a} where the separable operation $\{ \hat{M}_k^A \otimes  \hat{N}_k^B \}_{k\in {\cal S}}$ is applied to the state $\ket{\psi'}_{Ab}$ before the phase-flip channel $\Lambda^A_{|\braket{v_1}{v_0}|}$.

Let us consider the form of $\hat{N}_k^B$. 
From Eq.~(\ref{eq:m-k}), Eqs.~(\ref{eq:ld}) and (\ref{eq:ld'}) are reduced to
\begin{align}
\begin{cases}
\sqrt{\hat{\sigma}_k^{AB}} \ket{01}_{AB} =  \sqrt{\hat{\sigma}_k^{AB}} \ket{10}_{AB} =0 & (k\in {\cal S}_0), \\
\sqrt{\hat{\sigma}_k^{AB}} \ket{00}_{AB} =  \sqrt{\hat{\sigma}_k^{AB}} \ket{11}_{AB} =0 & (k\in {\cal S}_1).
\end{cases}
\label{eq:k}
\end{align}
From the definition (\ref{eq:defsigma}) of $\hat{\sigma}_k^{AB}$, 
$\hat{N}_k^B$ should satisfy 
\begin{equation}
\begin{cases}
\hat{N}_k^{B} \ket{u_0}_b =  n_k^{0} \ket{0}_B,\;  \hat{N}_k^{B} \ket{u_1}_b =   n_k^{1} \ket{1}_B & (k \in {\cal S}_0), \\
\hat{N}_k^{B} \ket{u_0}_b =  n_k^{0} \ket{1}_B,\;  \hat{N}_k^{B} \ket{u_1}_b =   n_k^{1} \ket{0}_B & (k \in {\cal S}_1) ,
\end{cases}
\label{eq:N0}
\end{equation}
with nonzero $n_k^{j}$.
Let $\{ {}_b\bra{\tilde{u}_i} \}_{i=0,1}$ be a dual basis in the Hilbert subspace spanned by ${}_b\bra{u_0}$ and ${}_b\bra{u_1}$ for the basis $\{\ket{u_i}_b\}_{i=0,1}$, which satisfies
\begin{equation}
\braket{\tilde{u}_i}{u_j}=\delta_{ij}. \label{eq:dual}
\end{equation}
By using this dual basis, $\hat{N}_k^{B}$ can be described by
\begin{align}
\hat{N}_k^{B}
= 
\begin{cases}
n_k^{0} \ket{0}_B {}_b \bra{\tilde{u}_0} + n_k^{1}  \ket{1}_B {}_b \bra{\tilde{u}_1} & (k\in {\cal S}_0), \\
n_k^{0} \ket{1}_B {}_b \bra{\tilde{u}_0} + n_k^{1}  \ket{0}_B {}_b \bra{\tilde{u}_1} & (k\in {\cal S}_1) .
\end{cases}
\label{eq:N}
\end{align}

As a result of Eqs.~(\ref{eq:m-k}) and (\ref{eq:N}), 
$\hat{\tau}_k^{AB}$ of Eq.~(\ref{eq:tau_k}) is described as $\hat{\tau}_k^{AB}=\Lambda^A_{|\braket{v_1}{v_0}|} (\ket{\psi'_k} \bra{\psi'_k}_{A B})$ with
\begin{multline}  
\ket{\psi'_k}_{AB}:= \frac{1}{\sqrt{p_k}} (\hat{M}_k^A \otimes \hat{N}_k^B) \ket{\psi'}_{A b}  \\
= \frac{1}{\sqrt{p_k}} \sum_{j=0,1} m_k^j n_k^j \sqrt{q_j} e^{i \Theta_j+i(-1)^j \varphi} (\hat{X}^A \hat{X}^{B})^l \ket{jj}_{AB} 
\label{eq:psi'_k}
\end{multline}
for $k \in {\cal S}_l$ ($l=0,1$),
where 
\begin{equation}
p_k = \sum_{j=0,1} q_j |m_k^{j} n_k^{j}|^2. \label{eq:prop}
\end{equation}
Since $\hat{\tau}_k^{AB}=\Lambda^A_{|\braket{v_1}{v_0}|}(\ket{\psi'_k}\bra{\psi'_k}_{AB})$ is in the standard form $\hat{\gamma}^{AB}(z_k, x_k)$, so is the state $\ket{\psi'_k}_{AB}$, namely, $\ket{\psi'_k}\bra{\psi'_k}_{AB}=\hat{\gamma}^{AB}(z_k',\sqrt{1-z_k'^2})$, where $z'_k$ is obtained from Eqs.~(\ref{eq:standard}) and (\ref{eq:psi'_k}) as
\begin{equation}
z_k'=\frac{2 \sqrt{q_0q_1}|m_k^0 m_k^1 n_k^0 n_k^1|}{p_k} . \label{eq:prosing}
\end{equation}
Considering the action of the phase-flip channel $\Lambda^A_{|\braket{v_1}{v_0}|}$,
we have
\begin{equation}
\begin{split}
&z_k=|\braket{v_1}{v_0}| z_k',  \\
&x_k=\sqrt{1-z_k'^2}. \label{eq:zx_k}
\end{split}
\end{equation}

Note that we cannot freely choose parameters $p_k$ and $z_k'$.
In particular, in order to make the operators $\{ \hat{M}_k^A \otimes \hat{N}_k^B \}$ achievable, the operators should satisfy 
Eq.~(\ref{eq:completeness-rel}). 
From Eqs.~(\ref{eq:m-k}), (\ref{eq:N}), and (\ref{eq:dual}), this condition is shown to be equivalent to
\begin{align}
\begin{split} 
\left(1-  \sum_{k \in {\cal S}} |m^0_k  n_k^0|^2 \right)^{1/2} \left(1-\sum_{k \in {\cal S}} |m^0_k  n_k^1|^2 \right)^{1/2} \ge |\braket{u_1}{u_0}|, \\
 \left(1-  \sum_{k \in {\cal S}} |m^1_k  n_k^0|^2 \right)^{1/2} \left(1-\sum_{k \in {\cal S}} |m^1_k  n_k^1|^2 \right)^{1/2} \ge |\braket{u_1}{u_0}|. \label{eq:normalize}
\end{split}
\end{align}
One way to derive these inequalities is to take a representation of $\hat{N}_k^{B\dag} \hat{N}^B_k$ with (orthonormal) cat states $\ket{c_\pm}_b:=(e^{-i\phi}\ket{u_0}_b+e^{i\phi}\ket{u_1}_b)/\sqrt{2(1\pm |\braket{u_1}{u_0}|)}$, where $2\phi:=\arg[\braket{u_1}{u_0}]$, and to notice that Eq.~(\ref{eq:completeness-rel}) means $\hat{1}^B - \sum_{k \in {\cal S}} m_k^j \hat{N}_k^{B\dag}\hat{N}_k^B \ge 0$ for $j=0,1$.

\subsection{An upper bound on separable operations}\label{se:2.3}

Let us derive an upper bound on the average overall yield $\bar{Y}$ defined in Eq.~(\ref{eq:performance-1}) assuming that $Y$ satisfies Eq.~(\ref{eq:mr}). From (\ref{eq:zx_k}), we have 
\begin{align}
\bar{Y} &= \sum_{k \in {\cal S}} p_k Y\left(|\braket{v_1}{v_0}| z_k',\sqrt{1-z_k'^2}\right) \nonumber \\
& \le Y\left(|\braket{v_1}{v_0}|, 0 \right) \sum_{k \in {\cal S}} p_k  z_k' . \label{eq:up}
\end{align}
On the other hand, since conditions of Eq.~(\ref{eq:normalize}) imply 
\begin{align}
\begin{split} 
\frac{1}{4}  \sum_{k \in {\cal S}}  \sum_{i,j=0,1} |m^i_k  n_k^j|^2   \le 1-  |\braket{u_1}{u_0}|, \label{eq:1-u}
\end{split}
\end{align}
we have
\begin{align}
\sum_{k\in {\cal S}} p_k z_k'& = 2 \sqrt{q_0 q_1}    \sum_{k\in {\cal S}}   |m^{0}_k n^{0}_k  m^{1}_k n^{1}_k | \nonumber \\
&\le \sum_{k\in {\cal S}}  |m^{0}_k n^{0}_k  m^{1}_k n^{1}_k |  
\le \frac{ 1}{4} \sum_{k\in {\cal S}} \sum_{i,j=0,1}  |m^{i}_k n^{j}_k |^2 \nonumber  \\
&\le  1-|\braket{u_1}{u_0}|. \label{eq:maxPF}
\end{align}
Therefore, substituting Eq.~(\ref{eq:maxPF}) for the bound of Eq.~(\ref{eq:up}),
we obtain an upper bound described by
\begin{align}
\bar{Y} \le Y(|\braket{v_1}{v_0}|,0)  (1-|\braket{u_1}{u_0}|) . \label{eq:y-upper''}
\end{align}
If we use Eq.~(\ref{tra}), this bound is rewritten as
\begin{align}
\bar{Y} &\le  Y(|\braket{u_1}{u_0}|^{\frac{1-T}{T}},0)  (1-|\braket{u_1}{u_0}|) ,\label{eq:y-upper}  
\end{align}
which gives an upper bound,
\begin{align}
\bar{Y}  
\le  \max_{0<u<1} Y(u^{\frac{1-T}{T}},0)  (1-u). \label{eq:y-upper'}
\end{align}

\subsection{An optimal protocol and the optimal performance}\label{se:2.4}

Conversely, here we show that the bound of Eq.~(\ref{eq:y-upper'}) is achievable by an entanglement generation protocol introduced in Ref.~\cite{A09}.
This protocol uses a dispersive Jaynes-Cummings Hamiltonian between a matter qubit and a coherent state leading to the assumption of $\ket{\alpha_0}_a=\ket{\alpha e^{i\theta/2}}_a$ and $\ket{\alpha_1}_a=\ket{\alpha e^{-i\theta/2}}_a$ with a constant $\theta>0$ in Eq.~(\ref{eq:V}), and it is regarded as a specific example of the protocol of Sec.~\ref{se:2.1}, based only on Bob's local operation composed of linear optical elements and ideal photon-number-resolving detectors.
The protocol provides \cite{A09} single-error-type entangled states when it succeeds, all of which can be transformed to $\hat{\gamma}^{AB}(2F-1,0)$
with fidelity 
\begin{equation}
F=\frac{1+u^{\frac{1-T}{T}}}{2},
\end{equation}
and the total success probability $P_{\rm s}=\sum_{k\in {\cal S}} p_k $ is
\begin{equation}
P_{\rm s} = 1-u,
\end{equation}
where $u:=|\braket{\alpha e^{i\theta/2}}{\alpha e^{-i\theta/2}}|^T$ is controllable by choosing $\alpha$.
Hence, with the entanglement generation protocol, the average overall yield $\bar{Y}$ is given by
\begin{align}
\bar{Y} &= \sum_{k \in {\cal S}} p_k Y(2F-1,0) = Y(2F-1,0) P_{\rm s} \nonumber \\
&= Y(u^\frac{1-T}{T},0) (1-u)  .
\end{align}
Since the parameter $u$ can be chosen freely in the protocol, the protocol can achieve 
\begin{equation}
\bar{Y}= \max_{0 < u < 1} Y \left(u^{\frac{1-T}{T}},0 \right)  (1-u), \label{eq:LOCC-max}
\end{equation}
which coincides with the upper bound (\ref{eq:y-upper'}).
Therefore, for the yield function $Y$ satisfying Eq.~(\ref{eq:mr}),
the entanglement generation protocol of Ref.~\cite{A09} is concluded to give the maximum yield of the single-error-type entanglement generation protocols.

\subsection{Comparison with the quantum/private capacity}\label{se:2.5}

 \begin{figure}[bt]
    \includegraphics[keepaspectratio=true,height=50mm]{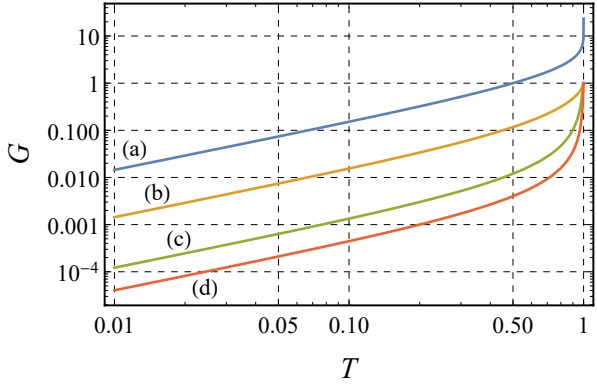}
  \caption{Performance of the point-to-point optimal supplier of single-error-type entanglement over coherent-state transmission over a lossy channel with the transmittance $T$. Curve~(a) describes the quantum/private capacity of the lossy channel as a reference. Curve~(b) describes $\bar{E}_{\rm D,max}$ which is the two-way distillable entanglement of the single-error-type entanglement generated by the optimal supplier, per channel use. Curves~(c) and (d) represent the success probabilities to obtain single-error-type entanglement with $F=99.4\%$ and with $F=99.8\%$ by using the optimal supplier, respectively.
  }
  \label{fig:P2P}
\end{figure}

To see how efficient the optimal protocol in Sec.~\ref{se:2.4} is, let us consider the asymptotic yield of an overall protocol, by assuming that the single-error-type entangled states successfully generated by the protocol are collectively input to the optimal two-way entanglement distillation protocol, implying the assumption of $Y=E_{\rm D}$. In this case, from Eqs.~(\ref{eq:E_D}) and (\ref{eq:LOCC-max}), the overall performance is 
\begin{equation}
\bar{E}_{\rm D,{\rm max}}:=\max_{0<u<1} (1-u)\left[ 1-h\left(\frac{1+u^\frac{1-T}{T}}{2} \right) \right].
\end{equation}
This quantity represents how many ebits are obtained per use of the entanglement generation protocol, i.e., per channel use, in an asymptotically faithful scenario. The overall yield $\bar{E}_{\rm D,{\rm max}}$ is plotted by the curve (b) of Fig.~\ref{fig:P2P}. On the other hand, the two-way quantum/private capacity of a pure-loss channel with the transmittance $T$ is $-\log_2 (1-T)$ \cite{PLOB17}, represented by the curve (a) of Fig.~\ref{fig:P2P}. We also describe the success probabilities to obtain single-error-type entanglement with $F=99.4\%$ and with $F=99.8\%$ by using the optimal entanglement generation in Sec.~\ref{se:2.4} as curves (c) and (d) of Fig.~\ref{fig:P2P}, respectively.
As shown in Fig.~\ref{fig:P2P}, the overall yield $\bar{E}_{\rm D,{\rm max}}$ based on the optimal two-way entanglement distillation (curve~(b)) is one order of magnitude less than the quantum/private capacity of the lossy channel (curve~(a)), and one order of magnitude better than direct generation of 99.4\%-fidelity entanglement with the optimal protocol in Sec.~\ref{se:2.4} (curve~(c)).

\section{Tight bound on single-error-type entanglement generation by three parties}\label{se:3}

In this section, we derive a tight bound on the entanglement generation protocols based on three parties.
We start by defining the protocols in Sec.~\ref{se:3.1}.
Using the results in Secs.~\ref{se:2.2} and \ref{se:2.3}, we derive an upper bound on the performance of protocols based on separable operations in Sec.~\ref{se:3.2}.
In Sec.~\ref{se:3.3}, we show that a protocol based on the remote nondestructive parity measurement of Ref.~\cite{ATKI10} achieves the upper bound.

\subsection{Single-error-type entanglement generation by three parties}\label{se:3.1}

\begin{figure}[b]
    \includegraphics[keepaspectratio=true,height=75mm]{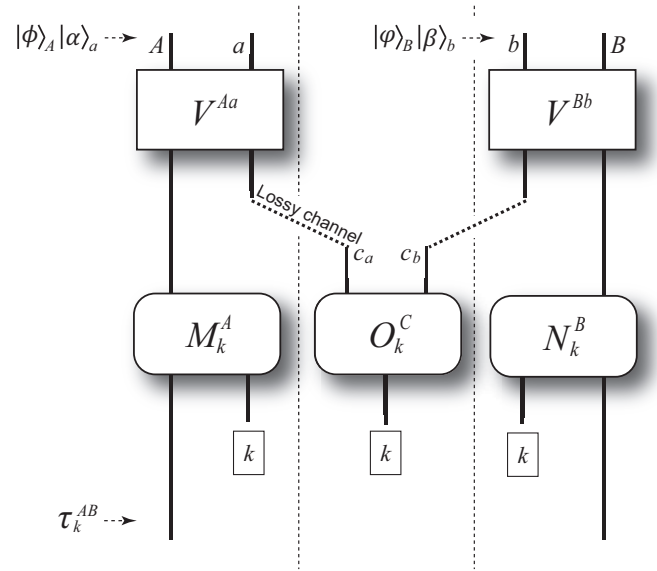}
  \caption{Scenario of single-error-type entanglement generation by three parties. $\ket{\phi}_A:=\sum_{j=0,1}  \sqrt{q_j^A} e^{i \Theta_j^A} \ket{j}_A$ and $\ket{\varphi}_B:=\sum_{j=0,1}  \sqrt{q_j^B} e^{i \Theta_j^B} \ket{j}_B$. An outcome $k$ corresponds to the application of a separable operator $\hat{M}_k^A \otimes \hat{N}_k^B \otimes {}_{C}\bra{O_k} $. If the final entanglement $\hat{\tau}_k^{AB}$ includes only one type of error
Alice, Bob and Claire may declare the success of the protocol.
   }
  \label{fig1b}
\end{figure}

To generate a single-error-type entangled state between Alice and Bob, 
they can ask another party called Claire for help.
In fact, single-error-type entanglement generation found in Ref.~\cite{ATKI10} adopts such a three-party protocol. 
This kind of protocol proceeds as follows (Fig.~\ref{fig1b}):
(i) Alice prepares qubit $A$ in her desired state $\ket{\phi}_A= \sum_{j=0,1} \sqrt{q_j^A}  e^{i \Theta_j^A} \ket{j}_A $ with real parameters $\Theta_j^A$, $q_j^A > 0$, and $\sum_j q_j^A=1$, and she makes it interact with a pulse in a coherent state $\ket{\alpha}_a=e^{-|\alpha|^2/2} e^{\alpha \hat{a}^\dag} \ket{0}_a$ via a unitary operation $\hat{V}^{Aa}$ of Eq.~(\ref{eq:V}).
(ii) Similarly, by using a unitary operation $\hat{V}^{Bb}$, Bob makes a pulse in coherent state $\ket{\beta}_b$ interact with
his qubit $B$ prepared in his desired state $\ket{\varphi}_B= \sum_{j=0,1} \sqrt{q_j^B} e^{i \Theta_j^B}  \ket{j}_B $.
(iii) Alice and Bob send the pulses $a$ and $b$ to Claire through lossy channels described by isometries $\hat{L}^{a\to c_aE_a}$ and $\hat{L}^{b\to c_bE_b}$,
\begin{equation}
\hat{L}^{x \to c_xE_x} \ket{\alpha}_x = \ket{ \sqrt{T_x} \alpha}_{c_x} \ket{\sqrt{1-T_x} \alpha}_{E_x}
\end{equation}
for $x=a,b$, respectively, 
where $c_x$ is the pulse at Claire's location, $0<T_x<1$ is the transmittance of the channel for pulse $x$, and $E_x$ is the environment.
At this point, Alice, Bob, and Claire share a quantum state $\ket{\xi }_{A c_a  E_a} \otimes \ket{\zeta }_{B c_b  E_b}$ with
\begin{align}
\begin{split}
&\ket{\xi }_{A c_a  E_a}= \sum_{j=0,1} \sqrt{q_j^A } e^{i \Theta_j^A}  \ket{j}_{A} \ket{\sqrt{T_a} \alpha_j}_{c_a} \ket{\sqrt{1-T_a} \alpha_j}_{E_a},  \\
&\ket{\zeta }_{B c_b  E_b}=\sum_{j=0,1} \sqrt{q_j^B}  e^{i \Theta_j^B } \ket{j}_{B} \ket{\sqrt{T_b} \beta_j}_{c_b} \ket{\sqrt{1-T_b} \beta_j}_{E_b} .\label{eq:initial-abc}
\end{split}
\end{align}
(iv) Alice, Bob, and Claire manipulate system $AB c_a c_b$ through LOCC, and may declare outcome $k$ with probability $p_k$ to herald the success of the generation of qubits $AB$ in a single-error-type entangled state $\hat{\tau}_k^{AB}$ in the form of Eq.~(\ref{eq:output-1}) for $k \in {\cal S}$.

This protocol is evaluated by the same way as in Sec.~\ref{se:2.1}, namely, by Eq.~(\ref{eq:performance-1}).

\subsection{An upper bound on single-error-type entanglement generation by three parties}\label{se:3.2}

Similar to Sec.~\ref{se:2.2}, we consider a separable operation $\{\hat{M}_k^{A} \otimes \hat{N}_k^B \otimes {}_{C}\bra{O_k} \}$, instead of the LOCC operation executed by Alice, Bob, and Claire at step (iv), based on the fact that separable operations compose a class of operations (strictly) larger than the set of LOCC operations.
Here, through finding the form of separable operators $\hat{M}_k^A \otimes \hat{N}_k^B \otimes {}_C\bra{O_k} $ that successfully return single-error-type entanglement $\hat{\tau}_k^{AB}$ in the form of Eq.~(\ref{eq:output-1}), 
we associate the three-party protocol with
a two-party protocol as in Fig.~\ref{fig2a}.

For simplicity, 
let us rewrite the states of Eq.~(\ref{eq:initial-abc}) as
\begin{align}
\begin{split}
& \ket{\xi}_{A c_a E_a} = \sum_{j=0,1} \sqrt{q_j^A} e^{i \Theta_j^A}  \ket{j}_A \ket{u_j^a}_{c_a}
  \ket{v_j^a}_{E_a} ,\\ 
& \ket{\zeta}_{B c_b E_b} = \sum_{j=0,1} \sqrt{q_j^B} e^{i \Theta_j^B}  \ket{j}_B \ket{u_j^b}_{c_b}
  \ket{v_j^b}_{E_b},
\label{eq:abe-abc}
\end{split}
\end{align}
where $0 < q_0^X < 1 $ and $q_0^X+q_1^X=1$ for $X=A,B$, and
\begin{equation}
\begin{split}
&1> |\braket{u_1^a}{u_0^a}|^{1-T_a} = |\braket{v_1^a}{v_0^a}|^{T_a}>0,\\
&1> |\braket{u_1^b}{u_0^b}|^{1-T_b} = |\braket{v_1^b}{v_0^b}|^{T_b}>0.
 \label{tra-abc}
\end{split}
\end{equation}
From Eqs.~(\ref{eq:abe-abc}) and (\ref{eq:phase-flip-ch}), we have 
\begin{equation}
\begin{split}
&{\rm Tr}_{E_a} [ \ket{\xi}\bra{\xi}_{A c_a E_a} ] = \Lambda^A_{|\braket{v_1^a}{v_0^a}|}(\ket{\xi'} \bra{\xi'}_{A c_a}), \\
&{\rm Tr}_{E_b} [ \ket{\zeta}\bra{\zeta}_{B c_b E_b} ] = \Lambda^B_{|\braket{v_1^b}{v_0^b}|} (\ket{\zeta'} \bra{\zeta'}_{B c_b}),
\end{split}
\end{equation}
where 
\begin{equation}
\begin{split}
&\ket{\xi'}_{A c_a}:= \sum_{j=0,1}  \sqrt{q_j^A} e^{i \Theta_j^A+i (-1)^j  \varphi_a }  \ket{j}_A \ket{u_j^a}_{c_a}, \\
&\ket{\zeta'}_{B c_b}:= \sum_{j=0,1}  \sqrt{q_j^B} e^{i \Theta_j^B + i (-1)^j  \varphi_b } \ket{j}_B \ket{u_j^b}_{c_b} 
\label{eq:xi'-abc} 
\end{split}
\end{equation}
with $2 \varphi_x:=\arg [\braket{v_1^x}{v_0^x}]$ for $x=a,b$. 
Hence, a separable operation $\{\hat{M}_k^A \otimes \hat{N}_k^B \otimes {}_C\bra{O_k} \}$ to the system $A B c_a c_b$ in state $\ket{\xi }_{A c_a  E_a} \otimes \ket{\zeta }_{B c_b  E_b}$ is equivalent to that in state $\Lambda^A_{|\braket{v_1^a}{v_0^a}|}(\ket{\xi'} \bra{\xi'}_{A c_a}) \otimes \Lambda^B_{|\braket{v_1^b}{v_0^b}|} (\ket{\zeta'} \bra{\zeta'}_{B c_b})$.

\begin{figure}[b]
    \includegraphics[keepaspectratio=true,height=58mm]{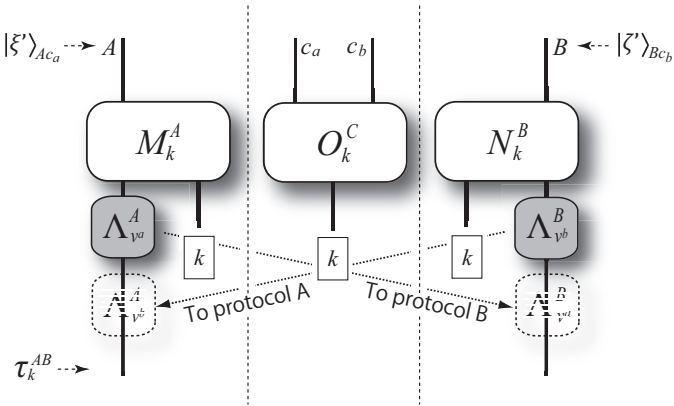}
  \caption{An imaginary protocol equivalent to the real protocol in Fig.~\ref{fig1b}. $\ket{\xi'}_{Ac_a}:=\sum_{j=0,1} \sqrt{q_j^A} e^{i \Theta_j^A+ i (-1)^j \varphi_a} \ket{j}_A \ket{u_j^a}_{c_a}$ and $\ket{\zeta'}_{Bc_b}:=\sum_{j=0,1} \sqrt{q_j^B} e^{i \Theta_j^B+ i (-1)^j \varphi_b} \ket{j}_B\ket{u_j^b}_{c_b}$. Channels $a \to c_a$ and $b \to c_b$ become ideal at the expense of the application of phase-flip channels $\Lambda^A_{v^a}$ with $v^a=|\braket{v_1^a}{v_0^a}|$ and $\Lambda^B_{v^b}$ with $v^b=|\braket{v_1^b}{v_0^b}|$, respectively. If the protocol returns success outcome $k$, the effect of phase-flip channel $\Lambda^B_{v^b} $ ($\Lambda^A_{v^a} $) is equivalent to that of $\Lambda^A_{v^b} $ ($\Lambda^B_{v^a} $). 
We define virtual protocol $A$ $(B)$ as the modified protocol where $\Lambda^B_{v^b} $ ($\Lambda^A_{v^a} $) is converted to $\Lambda^A_{v^b} $ ($\Lambda^B_{v^a} $).
   }
  \label{fig2b}
\end{figure}

Suppose that Alice, Bob, and Claire apply a separable operator $\hat{M}_k^A \otimes \hat{N}_k^B \otimes {}_C\bra{O_k} $ to the state $\Lambda^A_{|\braket{v_1^a}{v_0^a}|} (\ket{\xi'} \bra{\xi'}_{A c_a}) \otimes \Lambda^B_{|\braket{v_1^b}{v_0^b}|} (\ket{\zeta'} \bra{\zeta'}_{B c_b})$, and they return an
entangled state $\hat{\tau}_k^{AB}$ in the form (\ref{eq:standard}).
The separable operator $\hat{M}_k^A \otimes \hat{N}_k^B \otimes  {}_C\bra{O_k} $ must satisfy
\begin{equation} 
\begin{split} 
{}_{AB}\bra{01} \hat{M}_k^A \Lambda^A_{|\braket{v_1^a}{v_0^a}|}( \hat{\mu }_k^{AB} ) \hat{M}_k^{A\dag} \ket{01}_{AB} &=0,\\ \label{eq:simi}
{}_{AB}\bra{10} \hat{M}_k^A \Lambda^A_{|\braket{v_1^a}{v_0^a}|}( \hat{\mu }_k^{AB} ) \hat{M}_k^{A\dag} \ket{10}_{AB} &=0,
\end{split}
\end{equation}
with
\begin{multline}
\hat{\mu}_k^{AB}:=\hat{N}_k^B   {}_C\bra{O_k}  ( \ket{\xi'} \bra{\xi'}_{Ac_a} \\ 
\otimes \Lambda^B_{|\braket{v_1^b}{v_0^b}|}( \ket{\zeta'} \bra{\zeta'}_{Bc_b}))  \ket{O_k}_C \hat{N}_k^{B\dag}.
\end{multline}
Note that Eq.~(\ref{eq:simi}) is in the same form as Eqs.~(\ref{eq:single}) and (\ref{eq:single'}).
In addition,  $\hat{\mu }_k^{AB}$ is a positive operator and $0<|\braket{v_1^a}{v_0^a}|<1$.
Thus, similar to considerations from Eq.~(\ref{eq:single}) to Eq.~(\ref{eq:m-k}), 
$\hat{M}_k^{A}$ can be assumed to be in the form of 
\begin{align}
\hat{M}_k^A = 
\begin{cases} 
m_k^{0} \ket{0}\bra{0}_A +m_k^{1} \ket{1}\bra{1}_A & (k \in {\cal S}^A_0), \\
m_k^0 \ket{1}\bra{0}_A + m_k^1 \ket{0}\bra{1}_A & (k \in {\cal S}^A_1),
\end{cases}
\label{eq:aa}
\end{align}
with nonzero $m_k^{j}$, where ${\cal S}^A_0 $ and ${\cal S}^A_1$ are two disjoint subsets of set ${\cal S}$, i.e., satisfying ${\cal S}={\cal S}_0^A \cup {\cal S}_1^A$ and ${\cal S}_0^A \cap {\cal S}_1^A =\emptyset$, and $\hat{\mu}_k^{AB}$ satisfies
\begin{align}
\begin{cases}
\sqrt{\hat{\mu}_k^{AB}} \ket{01}_{AB} =  \sqrt{\hat{\mu}_k^{AB}} \ket{10}_{AB} =0 & (k \in {\cal S}^A_0), \\
\sqrt{\hat{\mu}_k^{AB}} \ket{00}_{AB} =  \sqrt{\hat{\mu}_k^{AB}} \ket{11}_{AB} =0 & (k \in {\cal S}^A_1).
\end{cases}
\label{eq:a}
\end{align}
This implies
\begin{equation}
\begin{split}
{}_{AB} \bra{01}  \hat{N}_k^B \Lambda^B_{|\braket{v_1^b}{v_0^b}|}( \hat{\nu }_k^{AB} ) \hat{N}_k^B \ket{01}_{AB}&=0, \label{eq:np}\\
{}_{AB} \bra{10}  \hat{N}_k^B \Lambda^B_{|\braket{v_1^b}{v_0^b}|}( \hat{\nu }_k^{AB} ) \hat{N}_k^B \ket{10}_{AB}&=0
\end{split}
\end{equation}
for any $k\in {\cal S}_0^A$ and
\begin{equation}
\begin{split}
{}_{AB} \bra{00}  \hat{N}_k^B \Lambda^B_{|\braket{v_1^b}{v_0^b}|}( \hat{\nu }_k^{AB} ) \hat{N}_k^B \ket{00}_{AB}&=0, \label{eq:np'}\\
{}_{AB} \bra{11}  \hat{N}_k^B \Lambda^B_{|\braket{v_1^b}{v_0^b}|}( \hat{\nu }_k^{AB} ) \hat{N}_k^B \ket{11}_{AB}&=0
\end{split}
\end{equation}
for any $k\in {\cal S}_1^A$, 
with
\begin{equation}
\hat{\nu}_k^{AB}:={}_C\bra{O_k} ( \ket{\xi'} \bra{\xi'}_{Ac_a} \otimes  \ket{\zeta'} \bra{\zeta'}_{Bc_b}) \ket{O_k}_C.
\end{equation}
Similar to considerations from Eq.~(\ref{eq:single}) to Eq.~(\ref{eq:m-k}), combined with $0<|\braket{v_1^b}{v_0^b}|<1$,
Eqs.~(\ref{eq:np}) and (\ref{eq:np'}) conclude that $\hat{N}_k^{B}$ can be assumed to be in the form of 
\begin{align}
\hat{N}_k^B = 
\begin{cases}
n_k^{0} \ket{0}\bra{0}_B +n_k^{1} \ket{1}\bra{1}_B & (k\in {\cal S}^B_0), \\
n_k^{0} \ket{1}\bra{0}_B +n_k^{1} \ket{0}\bra{1}_B & (k\in {\cal S}^B_1),
\end{cases}
\label{eq:a'}
\end{align}
where ${\cal S}_0^B$ and ${\cal S}_1^B$ are two disjoint subsets of set ${\cal S}$,
with nonzero $n_k^{j}$, and $\hat{\nu}_k^{AB}$ satisfies
\begin{equation}
\begin{cases}
\sqrt{\hat{\nu}_k^{AB}} \ket{01}_{AB} =  \sqrt{\hat{\nu}_k^{AB}} \ket{10}_{AB} =0  \\
\;\;\;\;\;\;\;\;\;\;\;\;\;\;\;\;\;\;\;\;\;\;\;\;\;\;\;\;\;\;\;\;\;\;\;\;\;\;(k\in \bigcup_{i=0,1 } {\cal S}_i^A \cap {\cal S}^B_i ),\\
\sqrt{\hat{\nu}_k^{AB}} \ket{00}_{AB} =  \sqrt{\hat{\nu}_k^{AB}} \ket{11}_{AB} =0  \\
\;\;\;\;\;\;\;\;\;\;\;\;\;\;\;\;\;\;\;\;\;\;\;\;\;\;\;\;\;\;\;\;\;\;\;\;\;\;(k\in \bigcup_{i=0,1 } {\cal S}_i^A \cap {\cal S}_{i\oplus 1}^B).
\end{cases}
\label{eq:nuortho}
\end{equation}
At this point, we have obtained two facts: (i) $\hat{M}_k^A \otimes \hat{N}_k^B $ commutes with the phase-flip channel $\Lambda^A_{|\braket{v_1^a}{v_0^a}|} \otimes \Lambda^B_{|\braket{v_1^b}{v_0^b}|}$; (ii) the range of $\hat{\nu}_k^{AB}$ is either the two-dimensional Hilbert subspace spanned by states $\{ \ket{00}_{AB}, \ket{11}_{AB} \}$ or that by states $\{\ket{01}_{AB},\ket{10}_{AB}\}$.
The fact (i) implies that the considered protocol is simulatable by a protocol of Fig.~\ref{fig2b} where the separable operation $\{ \hat{M}_k^A \otimes  \hat{N}_k^B \otimes {}_C\bra{O_k} \}$ is applied to the state $\ket{\xi'}_{Ac_a} \otimes \ket{\zeta'}_{Bc_b}$ before the phase-flip channel $\Lambda^A_{|\braket{v_1^a}{v_0^a}|} \otimes \Lambda^B_{|\braket{v_1^b}{v_0^b}|}$.
In addition, combined with Eqs.~(\ref{eq:aa}), (\ref{eq:a'}), and (\ref{eq:nuortho}), the fact (ii) indicates that 
$ \hat{M}_k^{A}  \hat{N}_k^{B} {}_C\bra{O_k} \ket{\xi'}_{Ac_a} \ket{\zeta'}_{Bc_b}$ belongs to the subspace spanned by states $\{ \ket{00}_{AB}, \ket{11}_{AB} \}$.
This implies that the effect of the phase-flip channel $\Lambda^B_{|\braket{v_1^b}{v_0^b}|}$ ($\Lambda^A_{|\braket{v_1^a}{v_0^a}|}$) on the entangled state $ \hat{M}_k^{A}  \hat{N}_k^{B} {}_C \bra{O_k} \ket{\xi'}_{Ac_a} \ket{\zeta'}_{Bc_b}$ is equivalent to that of a phase-flip channel $\Lambda^A_{|\braket{v_1^b}{v_0^b}|}$ ($\Lambda^B_{|\braket{v_1^a}{v_0^a}|}$).
Hence, in the success cases, the protocol works equivalently to a virtual protocol $A$ ($B$) in Fig.~\ref{fig2b} where Alice, Bob, and Claire prepare unnormalized state $ \hat{M}_k^{A}  \hat{N}_k^{B} {}_C \bra{O_k} \ket{\xi'}_{Ac_a} \ket{\zeta'}_{Bc_b}$ to be input into a series of phase-flip channels, $\Lambda_{|\braket{v_1^b}{v_0^b}|}^A \Lambda_{|\braket{v_1^a}{v_0^a}|}^A = \Lambda_{|\braket{v_1^a}{v_0^a}||\braket{v_1^b}{v_0^b}|}^A$ ($\Lambda_{|\braket{v_1^a}{v_0^a}|}^B \Lambda_{|\braket{v_1^b}{v_0^b}|}^B = \Lambda_{|\braket{v_1^a}{v_0^a}||\braket{v_1^b}{v_0^b}|}^B$).

Let us relate the virtual protocol $A$ in Fig.~\ref{fig2b} to the two-party protocol 
depicted in Fig.~\ref{fig2a} by regarding Claire and Bob in the former as a
single party, which is Bob in the latter.
Since Alice's operator $\hat{M}_k^A$ takes the same form (see Eqs.~(\ref{eq:aa}) and (\ref{eq:m-k})),
we notice that the protocol $A$ is a special case of the protocol in 
Fig.~\ref{fig2a} with the following substitutions:
\begin{equation}
\begin{split}
&q_j \mapsto q_j^A \\
&\Theta_j \mapsto \Theta_j^A \\
&\ket{u_j}_b \mapsto \ket{u_j^a}_{c_a} \\
&|\braket{v_1}{v_0}| \mapsto |\braket{v_1^a}{v_0^a}| |\braket{v_1^b}{v_0^b}| \\
&\hat{N}_k^B \mapsto ({}_C\bra{O_k}\otimes \hat{N}_k^B)\ket{\zeta'}_{Bc_b}.
\end{split}
\end{equation}
\if0
Let us relate the two-party protocol depicted in Fig.~\ref{fig2a} to
the virtual protocol $A$ with the phase-flip channel $\Lambda_{|\braket{v_1^a}{v_0^a}||\braket{v_1^b}{v_0^b}|}^A$, by regarding Alice versus Bob in the two-party protocol of Fig.~\ref{fig2a} as Alice versus Bob and Claire in the virtual protocol $A$. 
For given initial state $\ket{\xi'}_{A c_a}$ parametrized by  $\{q_j^A \}$ and $\{\ket{u_j^a}_{c_a} \}$, the initial state $\ket{\psi'}_{Ab}$ of Eq.~(\ref{eq:psi'}) is assumed to satisfy
\begin{equation}
\begin{split}
&q_j=q_j^A, \\
&\Theta_j=\Theta_j^A \\
&|\braket{u_1}{u_0}|=|\braket{u_1^a}{u_0^a}|,
\label{c-1}
\end{split}
\end{equation}
and it inevitably receives a phase-flip channel $\Lambda^A_{|\braket{v_1}{v_0}|}$ with 
\begin{equation}
|\braket{v_1}{v_0}|=|\braket{v_1^a}{v_0^a}| |\braket{v_1^b}{v_0^b}|. \label{c-2}
\end{equation}
Besides, Alice's operator $\hat{M}_k^A$ of Eq.~(\ref{eq:aa}) has already been in the form same as one of Eq.~(\ref{eq:m-k}), and Bob in the two-party protocol with Eqs.~(\ref{c-1}) and (\ref{c-2}) has less restrictions than  Bob and Claire in the virtual protocol $A$. 
\fi
Therefore, 
derivation of requisites starting from Eq. (\ref{eq:k}) to Eq. (\ref{eq:y-upper''})
is also applicable here under the above substitution. We thus obtain a bound,
\begin{equation}
\bar{Y} \le  Y(|\braket{v_1^a}{v_0^a}| |\braket{v_1^b}{v_0^b}|,0)  (1-|\braket{u_1^a}{u_0^a}|), \label{eq:y-upper-A}
\end{equation}
for the virtual protocol $A$. Similarly, by 
exchanging the roles of Alice and Bob in the above argument, we also obtain a bound,
\begin{align}
\bar{Y} \le Y(|\braket{v_1^a}{v_0^a}| |\braket{v_1^b}{v_0^b}|,0)  (1-|\braket{u_1^b}{u_0^b}|), \label{eq:y-upper-B}
\end{align}
for the virtual protocol $B$.
Let 
$u:=\max\{|\braket{u_1^a}{u_0^a}|,|\braket{u_1^b}{u_0^b}|\}$.
Since both inequalities (\ref{eq:y-upper-A}) and (\ref{eq:y-upper-B}) hold, use of Eqs.~(\ref{tra-abc}) and (\ref{eq:phase-flip-for-sing}) leads to
\begin{align}
\bar{Y} & \le Y(|\braket{v_1^a}{v_0^a}| |\braket{v_1^b}{v_0^b}|,0) (1-u) \nonumber \\
& \le  Y( u^{\frac{1-T_a}{T_a} + \frac{1-T_b}{T_b}} ,0) (1-u) .
\label{eq:y-upper-B'}
\end{align}
This gives an upper bound on $\bar{Y}$,
\begin{align}
\bar{Y}  
\le  \max_{0<u<1} Y( u^{\frac{1-T_a}{T_a} + \frac{1-T_b}{T_b}} ,0) (1-u) . \label{eq:y-upper-AB}
\end{align}

\subsection{An optimal protocol and the optimal performance for three-party protocols}\label{se:3.3}

Conversely, here we show that the bound of Eq.~(\ref{eq:y-upper-AB}) for protocols based on separable operations is achievable by a protocol based on the remote nondestructive parity measurement proposed in Ref.~\cite{ATKI10}.
Similar to the protocol \cite{A09} employed in Sec.~\ref{se:2.4}, this protocol uses unitary interaction of Eq.~(\ref{eq:V}) with the assumption of $\ket{\alpha_0}_a=\ket{\alpha e^{i\theta/2}}_a$ and $\ket{\alpha_1}_a=\ket{\alpha e^{-i\theta/2}}_a$ with a constant $\theta>0$, and it is based only on Claire's local operation composed of linear optical elements and ideal photon-number-resolving detectors.
The protocol provides \cite{ATKI10} single-error-type entangled states when it succeeds, all of which can be transformed to $\hat{\gamma}^{AB}(2F-1,0)$
with fidelity 
\begin{equation}
F=\frac{1+u^{\frac{1-T_a}{T_a} +\frac{1-T_b}{T_b}}}{2},
\end{equation}
and the total success probability $P_{\rm s}=\sum_{k\in {\cal S}} p_k$ is
\begin{equation}
P_{\rm s} = 1-u,
\end{equation}
where $u:= |\braket{\alpha e^{i\theta/2}}{\alpha e^{-i\theta/2}}|^{T_a}= |\braket{\beta e^{i\theta/2}}{\beta e^{-i\theta/2}}|^{T_b}$ is a parameter controllable by choosing $\alpha$ and $\beta$.
Therefore, with the entanglement generation protocol, the yield $\bar{Y}$ is
\begin{align}
\bar{Y} =& \sum_{k \in {\cal S}} p_k Y(2F-1,0) = Y(2F-1,0) P_{\rm s} \nonumber \\
=& Y(u^{\frac{1-T_a}{T_a} +\frac{1-T_b}{T_b}},0) (1-u)  .
\end{align}
Since $u$ can freely be chosen in the entanglement generation protocol of Ref.~\cite{ATKI10}, it can achieve the bound of (\ref{eq:y-upper-AB}), i.e.,
\begin{equation}
\bar{Y}= \max_{0 < u < 1} Y \left(u^{\frac{1-T_a}{T_a} +\frac{1-T_b}{T_b}},0 \right)  (1-u). \label{eq:LOCC-max-t}
\end{equation}

\subsection{Comparison with the quantum/private capacity}\label{se:3.4}

 \begin{figure}[tb]
    \includegraphics[keepaspectratio=true,height=50mm]{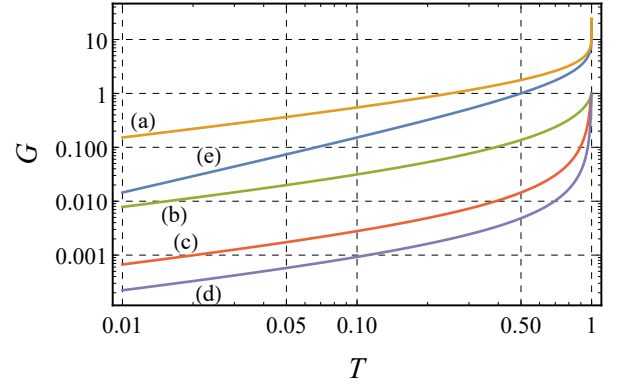}
  \caption{Performance of the optimal supplier of single-error-type entanglement with a middle measuring station. This middle station is connected with Alice and Bob by a lossy channel with transmittance $\sqrt{T}$, respectively (i.e., $T_a=T_b=\sqrt{T}$).
Curve~(a) describes the quantum/private capacity of this network. Curve~(b) describes $\bar{E}_{\rm D,max}$ which is the two-way distillable entanglement of the single-error-type entanglement generated by the optimal supplier, per use of the pair of channels between $AC$ and $CB$. Curves~(c) and (d) represent the success probabilities to obtain single-error-type entanglement with $F=99.4\%$ and with $F=99.8\%$ by using the optimal supplier, respectively. Curve~(e) describes the quantum/private capacity of a lossy channel with transmittance $T$ (which can directly connect Alice and Bob).
   }
  \label{fig:3P}
\end{figure}

Similar to Sec.~\ref{se:2.4},
to see how efficient the optimal protocol in Sec.~\ref{se:3.3} is, let us consider an asymptotic yield with $Y=E_{\rm D}$, by assuming that the single-error-type entangled states successfully generated by the protocol are collectively input to the optimal two-way entanglement distillation protocol. In this case, from Eqs.~(\ref{eq:E_D}) and (\ref{eq:LOCC-max-t}), the overall performance is 
\begin{equation}
\bar{E}_{\rm D,{\rm max}}:=\max_{0<u<1} (1-u)\left[ 1-h\left(\frac{1+u^{\frac{1-T_a}{T_a} +\frac{1-T_b}{T_b}}}{2} \right) \right].
\end{equation}
For simplicity, let $T_a=T_b=\sqrt{T}$, where $T$ represents the transmittance of a series of lossy channels between Alice and Claire and between Claire and Bob.
Then, the yield $\bar{E}_{\rm D,{\rm max}}$ is described by the curve (b) of Fig.~\ref{fig:3P}. On the other hand, the two-way quantum/private capacity (per use of the pair of channels between $A$ and $C$ and between $C$ and $B$) of the considered network is $-\log_2 (1-\sqrt{T})$ \cite{P19,R18,AK17,A21}, represented by the curve (a) of Fig.~\ref{fig:3P}. We also describe the success probabilities to obtain single-error-type entanglement with $F=99.4\%$ and with $F=99.8\%$ by using the optimal entanglement generation in Sec.~\ref{se:3.4} as curves (c) and (d) of Fig.~\ref{fig:3P}, respectively. As a reference, we also describe the quantum/private capacity $-\log_2 (1-T)$ of a lossy channel with transmittance $T$ (which can directly connect Alice and Bob) as curve~(e) of Fig.~\ref{fig:3P}.
As shown in Fig.~\ref{fig:3P}, the yield $\bar{E}_{\rm D,{\rm max}}$ based on the optimal two-way entanglement distillation (curve~(b)) is one order of magnitude less than the quantum/private capacity of the network (curve~(a)), and one order of magnitude better than direct generation of 99.4\%-fidelity entanglement with the RNPM protocol in Sec.~\ref{se:3.3}  (curve~(c)).

\section{Discussion}\label{se:4}

In this paper, we have identified entanglement generation based on the RNPM as the optimal supplier of single-error-type entanglement over coherent-state transmission, for arbitrary subsequent protocol with a jointly convex yield function satisfying Eq.~(\ref{eq:mr}). The condition (\ref{eq:mr}) is satisfied by typical yield functions, such as arbitrary convex functions of the singlet fraction and the distillable entanglement/key. 
If the distillable entanglement/key is adopted as a measure of the performance, 
its overall yield is only one order of magnitude less than the quantum/private capacity \cite{PLOB17,AK17,P19,R18} of the associated pure-loss bosonic channel network, and merely one order of magnitude better than direct generation of high-fidelity (in particular, 99.4\%-fidelity) entanglement with the RNPM, as represented by Fig.~\ref{fig:P2P} for the two-party protocol and by Fig.~\ref{fig:3P} for the three-party protocol. 
Considering that the overall yield cannot be achieved without the use of the optimal entanglement distillation in an asymptotic scenario, the latter gap implies that if entanglement generation protocol is efficient like one based on the RNPM, entanglement distillation protocol may not be necessary to achieve controlled-not operations in distributed quantum computation, in contrast to what one may infer from existing schemes \cite{FSG09,FYKI12,LB12,NLS13}. This suggests that performance of entanglement generation protocol affects the overall design of a distributed quantum computing architecture.
On the other hand,
considering that the quantum/private capacity \cite{PLOB17,AK17,P19,R18} of the associated pure-loss bosonic channel network is achieved with the use of two-mode infinitely squeezed vacuum states \cite{PGBL09},
the former gap implies that it is reasonable in practice to adopt protocol based on coherent-state transmission, like the RNPM protocol. Indeed, in the field of QKD, twin-field QKD protocol \cite{LYDS18,C19,CAL19,LL18,WYH18,MSK19,LNAKCR21,MZZ18} based on coherent-state encoding has already been identified as an important class of protocol to beat the repeaterless bounds \cite{TGW14,PLOB17} in a practical manner (see, e.g., \cite{CAL21}).

\section*{ACKNOWLEDGMENTS}
We would like to thank K.~Fujii, G.~Kato, W.~J.~Munro, N.~Sota and H.~Takeda for valuable discussions. 
We acknowledge the support from Moonshot R\&D, JST JPMJMS2061, in part, from PREST, JST JP-MJPR1861, and from JSPS KAKENHI 21H05183 JP.

\end{document}